\documentclass[12pt,preprint]{aastex}

\begin{document}

\title{Temporal and Spectral Characteristics of Short Bursts
      from the Soft Gamma Repeaters 1806-20 and 1900+14}

\author{Ersin {G\"o\u{g}\"u\c{s}}\altaffilmark{1,2}, 
Chryssa Kouveliotou\altaffilmark{2,3},
Peter M. Woods\altaffilmark{2,3}, 
Christopher Thompson\altaffilmark{4},
Robert C. Duncan\altaffilmark{5},
Michael S. Briggs\altaffilmark{1,2}}

\altaffiltext{1}{Department of Physics, University of Alabama in Huntsville,
       Huntsville, AL 35899} 
\altaffiltext{2}{NASA Marshall Space Flight Center, NSSTC SD-50, 
       320 Sparkman Dr., Huntsville, AL 35805}
\altaffiltext{3}{Universities Space Research Association}
\altaffiltext{4}{Canadian Institute for Theoretical Astrophysics,
       60 St. George St., Toronto, ON, Canada M5S 3H8}
\altaffiltext{5}{Department of Astronomy, University of Texas, RLM 15.308, 
       Austin, TX 78712-1083}

\authoremail{Ersin.Gogus@msfc.nasa.gov}

\begin{abstract}

We study the temporal and coarse spectral properties of 268 bursts from 
SGR 1806$-$20 and 679
bursts from SGR 1900+14, all observed with the Rossi X-Ray Timing
Explorer/Proportional Counter Array. Hardness ratios and temporal parameters, 
such as T$_{90}$
durations and $\tau$$_{90}$ emission times are determined for these bursts.
We find a lognormal distribution of burst durations, 
ranging over more than two orders of magnitude:
T$_{90}$ $\sim$ 10$^{-2}$ to $\gtrsim$ 1 s, with a peak at $\sim$ 0.1 s.
The burst light curves tend to be asymmetrical, with more than half of
all events showing rise times t$_{\rm r}$ $<$ 0.3 T$_{90}$.
We find that there exists a correlation between the duration
and fluence of bursts from both sources. We also find a
significant anti-correlation between hardness ratio and fluence for SGR
1806$-$20 bursts and a marginal anti-correlation for SGR 1900+14 events. 
Finally, we discuss possible physical implications of these results 
within the framework of the magnetar model.

\end{abstract}


\section{Introduction}

Soft gamma repeaters (SGRs) are a small class of objects that are
characterized by brief and very intense
bursts of soft gamma-rays and hard X-rays. 
They are distinguished from classical gamma-ray bursts (GRBs)
by their repeated periods intense activity, during which 
dozens of bursts with energies approaching 10$^{41}$ ergs are recorded.
SGR bursts have significantly softer
spectra than classical GRBs; the former are being well fit by an 
optically-thin thermal bremsstrahlung model with 
temperatures $kT = 20-40$ keV.  Two SGRs (0526-66 and 1900+14)
have emitted one giant flare each: events that are much more energetic
($E \sim 10^{44-45}$ erg) and contain a very hard spectral 
component within the first $\sim$ 1 s (Mazets et al. 1979; Cline et al. 1981;
Hurley et al. 1999; Mazets et al. 1999; Feroci et al. 1999).
For a review of the burst and 
persistent emission properties of SGRs, see Hurley (2000).

In 1992, it was suggested that SGRs are strongly magnetized 
($B \gtrsim 10^{14}$ G) neutron stars, or
magnetars (Duncan \& Thompson 1992, see also Pac{\`z}ynski 1992).
This model suggests crustquakes as a plausible trigger for the short
SGR bursts (as well as the giant flares):  a sudden fracture
of the rigid neutron star crust, driven by the build-up of 
crustal stresses as the strong magnetic field gradually 
diffuses through the dense stellar matter
(Thompson \& Duncan 1995 [hereafter TD95]; 1996).  The motion of the crust 
shears and twists the external magnetic field, and in the process
releases both elastic and magnetic energy.  

Cheng et al. (1996) studied a set of 111 SGR 1806$-$20 bursts
(detected with the International Cometary Explorer, ICE) and determined that
some of their properties, such as size and cumulative waiting-time distributions
are similar to those of earthquakes.
Recently, G\"o\u{g}\"u\c{s} et al. (2000) confirmed these
similarities for SGR 1806$-$20 using a larger sample of 401 bursts 
(111 detected with the Burst and Transient Source Experiment, BATSE,
aboard the Compton Gamma Ray Observatory, CGRO; 290 with the
the Proportional Counter Array, PCA, on the Rossi X-ray Timing Explorer, RXTE).
The same similarities were found for SGR 1900+14 when we analyzed
a sample of 1024 bursts (187 observed with CGRO/BATSE; 837 with the RXTE/PCA) 
(G\"o\u{g}\"u\c{s} et al. 1999).
Furthermore, we reported evidence for a correlation between duration
and fluence for the SGR 1900+14 bursts, similar to the correlation seen 
between
the duration of strong ground motion at short distances from an earthquake 
region and the seismic moment ($\propto$ energy) of an earthquake
(Lay \& Wallace 1995). This similarity strongly suggests that SGR bursts, like
earthquakes, may be manifestations of self-organized critical systems 
(Bak, Tang \& Wiesenfeld 1988), lending support to the hypothesis
that SGR bursts involve the release of some form of stored potential
energy.

Statistical studies such as the above provide important
clues not only about the mechanism by which energy is injected
during an SGR burst, but also the mechanism by which it is radiated.
For example, in the magnetar model the stored potential energy
may be predominantly magnetic, and the electrical currents
induced by crustal fractures may be
strongly localized over small patches of the star's surface, 
somewhat as they are in solar
flares.  In such a situation, the energy density in the non-potential
magnetic field is high enough that excitation of high
frequency Alfv\'en motions leads to rapid damping and the creation
of a trapped fireball (through a cascade to high wavenumber:
Thompson \& Blaes 1998).  By contrast, if energy is injected more
globally, then an approximate balance between injection and radiation
will occur in the bursts of luminosity less than $\sim 10^{42}$ erg
s$^{-1}$ (Thompson et al. 2000).

Here, we investigate detailed temporal characteristics of a large subset 
of SGR 1806$-$20
and SGR 1900+14 bursts observed with the RXTE/PCA during their burst active
periods in 1996 and 1998, respectively. 
We apply to the SGR bursts some temporal analysis methods that were
originally developed for the study of GRBs.
We also study the spectral variations of the bursts as a function of
the burst fluence and duration. In \S~ 2, we briefly review the RXTE/PCA
observations. Section 3 describes the data analysis techniques and our 
results are discussed in \S~ 4.

\section{Observations}

The PCA instrument (Jahoda et al. 1996)
consists of five Xe proportional counter units (PCUs) sensitive to
energies between 2$-$60 keV with 18\% energy resolution at 6 keV. The PCA has a 
total effective area of $\sim$ 6700 cm$^{2}$. 

{\bf SGR 1806$-$20 : }
The RXTE/PCA observations of SGR 1806$-$20 were performed during a burst active
period of the source in 1996 (between November 5 and 18) for a total effective 
exposure time
of 136.8 ks. Using the burst search procedure described in 
G\"o\u{g}\"u\c{s} et al. (2000), we have identified 290 bursts from the source.
The number of integrated counts (2$-$60 keV) for these bursts range 
between 22 and 34550. Using the count-to-energy conversion factor
given in G\"o\u{g}\"u\c{s} et al. (2000) this corresponds to burst fluences
between $1.2 \times 10^{-10}$ and $1.9 \times 10^{-7}$ ergs cm$^{-2}$ 
(E $>$ 25 keV).
Assuming isotropic burst emission, the corresponding energy range is
$3.0 \times 10^{36}$ -- $4.9 \times 10^{39}$ ergs, for a distance to 
SGR 1806$-$20 of 14.5 kpc (Corbel et al. 1997).
To facilitate our analysis, we chose events that were clustered together during
two very active epochs of the source, resulting in 268 events recorded on
5 (MJD 50392) and 18 (MJD 50405) November 1996.

{\bf SGR 1900+14 : }
RXTE observations of SGR 1900+14 took place
between 1998 June 2 and December 21, for a total effective exposure time of
224.1 ks. Using the same burst search algorithm as before, we identified a 
total of 837 bursts from the source with integrated counts
ranging between 22 and 60550. The fluences of these bursts range 
from $1.2 \times 10^{-10}$ to $3.3 \times 10^{-7}$ ergs cm$^{-2}$ (E $>$ 25 keV) 
corresponding to an energy range of $7 \times 10^{35}$ -- $2 \times 10^{39}$ 
ergs (assuming isotropic emission at 7 kpc [Vasisht et al.~1994]). 
Similarly, we selected 679 events which occurred during a very active period of
the source between 1998 August 29 (MJD 51054) and September 2 (MJD 51058).

\section{Data Analysis and Results}

\subsection{Duration (T$_{90}$) estimates}

Originally defined for cosmic GRBs, the T$_{90}$ duration of a burst is the time
during which 90\% of the total (background-subtracted) burst counts have been
accumulated since the burst trigger (Kouveliotou et al. 1993). 
We calculated the 
T$_{90}$ duration of SGR bursts using event-mode PCA data (2-60 keV)
with 1/1024 s time resolution. For each burst we collected the cumulative counts
for an 8 s continuous stretch of data starting 4 s before its peak, t$_{\rm p}$.
We then fit the cumulative count distribution 
between two user-selected background intervals\setcounter{footnote}{5}
\footnote{Approximately equal background intervals were selected before and
after the burst, except in cases where separation between bursts was too small;
there we chose smaller, but significant, post/pre burst intervals.} 
with a first order polynomial
plus a step function (for the burst) and subtracted the background 
counts. By using a first order polynomial to fit the cumulative counts, we 
assume the PCA background remains flat over each 8 s segment. The resulting
height of the step function gives the total burst counts. Figure 1 depicts the
steps of the T$_{90}$ estimate procedure for one of the SGR 1900+14 bursts.

As the selected bursts occurred during extremely active periods 
(for both sources), there were quite a number of
cases where bursts were clustered very close together.
During these active episodes, many crustal sites on the neutron star
may be active, releasing stored potential energy in the form of bursts
with a large variety of time profiles.
Hence, it is important to distinguish single pulse events from events 
with multiple peaks (which may involve multiple fracture sites). 
To do this, we applied arbritrary but consistent criteria to our data.
We classified an event as {\it multi-peaked} if the count rate at any local
minimum (in 7 ms time bins) is less than half the maximum value attained 
subsequently in the burst (see Figure 2, middle plots).  
Otherwise the event was classified as {\it single-peaked} (Figure 2, 
top plots).  When the count rate dropped to the noise level between peaks, 
the event was classified as a single, multi-peaked burst if and only if the 
time between peaks was less than a quarter of the neutron star rotation 
period ( 1.3 s for SGR 1900+14, 1.9 s for SGR 1806-20; see Figure 2, bottom 
plots).  We found that 262 of the 679 bursts from SGR 1900+14, and 113 of 
the 268 SGR 1806-20 bursts were multi-peaked by these criteria.   
Note that the total counts per burst (or "count fluences") were found
to span nearly the same range in the two classes of bursts, in both 
sources.  This suggests that the light curve peak structure 
does not correlate strongly with total energy.

In the context of cosmic GRBs, the value of T$_{90}$ can be systematically 
underestimated in faint events due to low signal-to-noise (Koshut et al. 1996). 
We investigated this effect in our SGR analysis 
using extensive numerical simulations. We created three time profiles based 
upon the observed time profiles; a single two-sided 
Gaussian (with right-width twice the left width), two two-sided Gaussians whose 
peaks are separated by 0.5 s and the peak rate of the second pulse is 0.6 of 
the first one,
and two two-sided Gaussians with peak separation of 1 s. For each profile, we
varied the peak rates (eight values between 2600 counts s$^{-1}$ and 55000 
counts s$^{-1}$)
and the width of the Gaussians (eight values from 12 ms to 120 ms). 
We determined 
the T$_{90}$ duration for each combination before adding noise. Then for each
combination, we generated 800 realizations including Poisson noise and
determined the respective T$_{90}$ values for these simulations. We found
that the fractional difference, FD = 1 $-$ (T$_{90,\rm s}$ / T$_{90,\rm a}$)
between the actual value, T$_{90,\rm a}$, 
(in the absence of noise) and the simulated one, T$_{90,\rm s}$, has a 
strong dependence on the peak rate (i.e. signal-to-noise ratio), 
such that FD can be as high as 0.22 at count
rates of 2600 counts s$^{-1}$. 
It is, however, less than 0.05 for count rates greater
than 4800 counts s$^{-1}$. The variation of FD with respect to 
the pulse profile is insignificant.   
Using the results of our simulations we obtained a 
T$_{90}$ correction as a function of peak rate.
We estimated the peak
rate of each burst using a box-car averaging technique with a box width of 
1/512 s.
We then corrected T$_{90}$ values for the S/N effect using our correction 
function and these peak rates.

Our final data set thus comprises 455 SGR 1900+14 bursts with corrected
$T_{90}$ durations between 9 ms and 2.36 ms (Figure 3, solid histogram).
We were unable to determine statistically significant $T_{90}$ durations 
for 187 SGR 1900+14 bursts because there were less than $\sim 40$
counts per event.
The solid curve in Figure 3 is a best fit log-Gaussian 
function to all T$_{90}$ values which peaks at 93.9 $\pm$ 0.2 ms 
($\sigma$ = 0.35 $\pm$ 0.01; where $\sigma$ is the width of the distribution 
in decades). The dashed histogram in Figure 3 is the distribution of 
T$_{90}$ values of single pulse bursts whose log-Gaussian mean is 
46.7 $\pm$ 0.1 ms ($\sigma$ = 0.21 $\pm$ 0.01). Also in Figure 3, 
the dash-dot histogram displays the T$_{90}$ distribution of multi-peaked
bursts and it peaks at 148.9 $\pm$ 0.2 ms ($\sigma$ = 0.26 $\pm$ 0.02).
For SGR 1806$-$20 bursts, we determined corrected T$_{90}$ values of 190 bursts
which range between 16 ms and 1.82 s (49 bursts were too weak to determine 
their T$_{90}$ values). 
The distribution of all T$_{90}$ values
is shown in Figure 4 (solid histogram). A log-Gaussian fit to this distribution
yields a peak at 161.8 $\pm$ 0.2 ms ($\sigma$ = 0.34 $\pm$ 0.02).
Similarly, a log-Gaussian fit to the
T$_{90}$ distribution of the single pulse bursts (dashed lines in Figure 4) 
peaks at 88.1 $\pm$ 0.1 ms ($\sigma$ = 0.19 $\pm$ 0.03) and a fit to the 
multi-peaked bursts (dash-dot lines in Figure 4) yields a peak at
229.9 $\pm$ 0.3 ms ($\sigma$ = 0.32 $\pm$ 0.03).

In order to quantify the time profile symmetry of SGR bursts we determined the 
ratio of their rise times (t$_{r}$, i.e. the interval between T$_{90}$ start 
time and peak time t$_{p}$) to T$_{90}$ durations of the single pulse events. 
In Figure 5, we plot the distributions of these ratios for bursts with single 
pulse (solid histogram) and milti-peaks (dashed histogram) from 
SGR 1900+14 (left) and those from SGR 1806$-$20 (right). 
The majority of single pulse events from both sources have t$_{r}$ / T$_{90}$
values less than 0.3, showing that SGR bursts decay slower than they rise, 
or in other words that SGR pulse profiles are asymmetric. We find that the 
average values os the t$_{r}$ / T$_{90}$ ratios are 0.29 and 0.27 for 
SGR 1900+14 and 
SGR 1806$-$20, respectively. Remarkably coincident average values, along with
very similar distributions of t$_{r}$ / T$_{90}$ ratios of both sources suggest
a similarity of the asymmetry in the temporal profiles of SGR bursts.

\subsection{Emission time ($\tau$$_{90}$) estimates}

Emission time, $\tau$$_{N}$, was introduced for cosmic GRBs as a 
complimentary temporal parameter to T$_{90}$
(Mitrofanov et al. 1999). Emission time is the time
over which a fixed percentage, N\%, of the total burst emission was
recorded, starting from the peak of the event and moving downward in flux.
For each burst, we determined the average background level from the background
intervals used in the T$_{90}$ estimating procedure. The
background-subtracted count bins within the burst interval (i.e. from
the end of pre-burst background range to the beginning of post-burst background 
range) were then ordered by decreasing count rate. 
Starting with the highest count rate bin (i.e. the peak), 
we added the counts of each successively weaker bin until 90\%
of the total burst counts were accumulated. 
The $\tau$$_{90}$ emission time of the
burst is then the total time spanned by the accumulated bins. 

As with the T$_{90}$ parameter, when the S/N ratio of the burst is
low, the measured emission time can be systematically smaller than the 
actual value (Mitrofanov et al. 1999). In order to correct for this
systematic error, we constructed numerous simulated profiles identical to those 
described in the previous section.
We found that, similar to T$_{90}$ estimates, the FD 
between the actual and simulated $\tau$$_{90}$ emission times is strongly
dependent on the peak rate. FD is $\sim$ 0.26 at count rates of 2600 
counts s$^{-1}$
and less than 0.06 for count rates greater than 4800 counts s$^{-1}$.
Similar to the T$_{90}$ measurement, this effect is only weakly dependent on the
temporal profile of the burst.
Using our $\tau$$_{90}$ correction function and the peak rates,
we obtained the corrected $\tau$$_{90}$ emission times. The corrected values
range between 4.6 and 412.8 ms for SGR 1900+14 bursts, and between 
4.8 and 559.2 ms for SGR 1806$-$20 bursts.
In Figure 6, we show the distributions of $\tau$$_{90}$ emission times for
SGR 1900+14 ({\it left}) and SGR 1806$-$20 ({\it right}). The dashed lines in
both plots are the best fit log-Gaussian curves which peak at 49.6 $\pm$ 0.1 ms
($\sigma$ = 0.28 $\pm$ 0.01) for SGR 1900+14 events, and at 82.3 $\pm$ 0.1 ms 
($\sigma$ = 0.32 $\pm$ 0.02) for SGR 1806$-$20 events.
The $\tau$$_{90}$ emission times of bursts from both sources are on average 
shorter than T$_{90}$ durations. Note that the log-Gaussian
mean values of T$_{90}$ distributions of {\it single} pulse bursts from both 
SGRs are quite similar to those of the $\tau$$_{90}$ distributions.

\subsection{Duty cycles ($\delta$$_{90}$)}

As shown in the previous section, the $\tau$$_{90}$ values of bursts are in 
general smaller than their T$_{90}$ values. 
The reason is that $\tau$$_{90}$ is not a
measure of the actual duration of the burst when the morphology of the burst is
complex (e.g. multiple peaks). 
Mitrofanov et al. (1999) suggested that the ratio $\tau$$_{90}$/T$_{90}$ 
can then be used to describe a duty cycle, $\delta$$_{90}$. 
We, therefore, determined the $\delta$$_{90}$ parameters for all bursts of both
sources (Figure 7, {\it left} for SGR 1900+14 bursts, {\it right} for SGR
1806$-$20 bursts). Both distributions peak at about $\delta$$_{90}$ $\sim$ 0.45
and their overall shapes are very similar, indicating a strong similarity of the
{\it type} of burst emission in the two sources, although their overall duration
distribution differ.

\subsection{Duration -- Fluence -- Hardness correlations}

Gutenberg and Richter (1956) showed that there exists a power law 
relation between the magnitude (which is related to the total energy involved) 
of earthquakes and the durations of strong ground-shaking
at short distances from an earthquake epicenter. We investigated whether a 
similar
correlation exists in SGR events using their total burst counts and their 
T$_{90}$ durations (as estimated in \S 3.1), 
for both SGR 1806$-$20 and SGR 1900+14.
For each set of bursts, we grouped the T$_{90}$ values
into logarithmically spaced bins and determined the weighted mean 
value of total burst counts and T$_{90}$ durations for each bin. 
In Figure 8, we show the plot of integral counts vs T$_{90}$ durations
of SGR 1806$-$20 bursts. The crosses in this figure are the weighted means of
each parameter; the errors on the mean-T$_{90}$ values denote the range of 
each bin while the errors on the mean counts are 
due to sample variance. The dark points are the
individual measurements of the burst counts and T$_{90}$ values of single pulse
events and the gray circles are those of multi-peaked bursts.
Figure 8 shows that the integral counts and durations of SGR 1806$-$20 bursts 
are well correlated. To quantify this correlation, we determined the 
Spearman's rank order correlation coefficient, $\rho$ = 0.91, and the
probability of getting this value from a random data set, P=3.4 $\times$
10$^{-4}$. We further fit a power law model to the mean values (crosses) of the
data using the least squares technique, which yields a power law index of 
1.05 $\pm$ 0.16.

Similarly, Figure 9 shows that the integral counts and T$_{90}$ durations of 
SGR 1900+14 bursts are also correlated, having 
$\rho$ = 0.89 and P = 8.6 $\times$ 10$^{-4}$. A power law fit
to the mean values of the data yields an index of 0.91 $\pm$ 0.07. The asterisk
shown in the upper right portion of Figure 9 indicates where the precursor of
the August 29 burst falls. This event was exceptionally long, bright and 
resembled 
the August 27 giant flare in various ways (Ibrahim et al. 2000).
A spectral line at 6.4 keV was reported during the precursor of event of
August 29 burst with $\sim$ 4 $\sigma$ significance 
(Strohmayer \& Ibrahim 2000). 
If spectral lines are event intensity dependent, there are very few events
in Figure 9 during which we would expect to see any lines. The detailed 
spectral analysis of all SGR bursts is underway.

In conclusion we find a good correlation between the duration and total counts
(energy) of SGR bursts, quite similar to the one established for earthquakes.
It is noteworthy that the burst count fluences of single pulse events (from both 
systems) span a range almost as wide as that of fluences of multi-peaked 
bursts.   

In order to investigate burst spectral variations versus 
fluence, we calculated an event hardness ratio 
defined as the ratio of the total burst counts observed between 10$-$60 keV 
to those between 2 $-$ 10 keV.
Bursts with total counts less than $\sim$ 50 yielded
statistically insignificant hardness ratios ($<$ 3 $\sigma$), and therefore,
were excluded from our analysis. 
For SGR 1806$-$20 bursts, we have 159 events with hardness ratios
measured to $>$ 3 $\sigma$ accuracy.
We divided the total counts of these events into logarithmically
spaced bins and determined the weighted mean
hardness ratio for each group. Figure 10a shows that the SGR 1806$-$20
burst hardness ratios are anti-correlated with fluence 
($\rho$=$-$0.96, P=2.6 $\times$ 10$^{-4}$).  
For 385 SGR 1900+14 bursts, Figure 10d shows a marginal 
anti-correlation between hardness and fluence
($\rho$ = $-$0.89, P=4.6 $\times$ 10$^{-3}$).
Interestingly, we find that although the hardness $-$ fluence anti-correlation
is evident in both sources, the SGR 1806$-$20 events are {\it overall} harder
than the SGR 1900+14 ones. Our energy selection optimizes the PCA energy
response, so that we are not affected by instrumental biases.

Fenimore, Laros \& Ulmer (1994) have performed a similar study of hardness 
ratio vs
fluence for 95 SGR 1806-20 events detected with ICE. We cannot directly compare
our results with this study for two reasons. The energy ranges over which 
hardness ratios are computed differ significantly; while we use 
(10 $-$ 60 keV)/(2 $-$ 10 keV), Fenimore et al. (1994) use 
(43.2 $-$ 77.5 keV)/(25.9 $-$ 43.2 keV). The PCA sensitivity drops significantly
beyond $~$ 25 keV, limiting the possibility of comparison.
Furthermore, our fluence range ends at $~2 \times 10^{-7}$ ergs cm$^{-2}$, 
just below the range described in Figure 1 of Fenimore et al. (1994).
However, since Fenimore et al. report a constant trend above 
$10^{-7}$ ergs cm$^{-2}$, one could postulate that the hardness
ratio vs fluence trend we report, levels off at higher fluences (energies).

We further investigated the hardness $-$ fluence trend for burst morphology
sub-sets, namely for single pulse and multi-peaked burst groups for each 
source. 
Figure 10b exhibits hardness $-$ fluence plot of single pulse SGR 1806$-$20 
events and Figure 10c shows that of multi-peaked bursts.
We see that both sets display spectral softening as the burst count fluence
increases. Similarly SGR 1900+14 events are shown in Figure 10e and f.
For this source, the hardness $-$ fluence anti-correlation is significant only
for single pulse bursts. 
It is important to note that the peak rates of most of the 
highest fluence bursts reach $\gtrsim$ 10$^5$ counts s$^{-1}$ (on 1/1024 s 
time scale) around which the PCA pulse pileup may become important. This effect 
can
artificially harden the observed count spectrum at these rates. The spectral
hardening or leveling off seen in the last bins of the plots of Figure 10
may well be due to the pulse pileup effect and should not be considered as
intrinsic source property.
 
We next investigated the relationship between the hardness and duration of SGR 
bursts. Although we find hardness $-$ fluence anti-correlations and
fluence $-$ duration correlations, both SGR 1806$-$20 (Figure 11, squares) and
SGR 1900+14 (Figure 11, diamonds) burst hardness ratios are independent of 
the event durations.

\section{Discussion}

Our study demonstrates that unlike the T$_{90}$ duration distribution of cosmic
GRBs which shows a bi-modal trend with peaks at $\sim$ 0.31 s and $\sim$ 37 s
(Kouveliotou et al. 1993; Paciesas et al. 1999), the T$_{90}$ distribution of 
SGR bursts displays a single peak which varies for each source: $\sim$ 93 ms
and $\sim$ 162 ms for SGR 1900+14 and SGR 1806$-$20, respectively. 

We find that the ${\rm T}_{90}$ durations of single-pulse bursts
from both SGRs form narrow distributions (compared
to those of multi-peaked events) which peak at $\sim 47$ ms
and $\sim 88$ ms for SGR 1900+14 and SGR 1806-20, respectively.
These bright, hard, and short bursts are almost certainly powered
by a sudden disturbance of the rigid neutron star crust, which
transmits energy to the magnetosphere.  Magnetic stresses,
which force the star between distinct metastable equilibria,
provide the most plausible source of energy for the giant flares,
and by inference for the short bursts (TD95; Thompson \& Duncan 2001).

The duration of the multi-peaked bursts is clearly fixed by the 
time between successive releases of energy.  The widths of the
single pulse bursts could, in principle, be limited either
by the rate of release of energy from the initial reservoir 
or, alternatively, by the time for the released energy to be
converted to radiation through some intermediate reservoir.   There have
been various suggestions for such an intermediate storage mechanism:
a hot fireball that is confined on closed magnetic field lines 
(TD95);   a region of strong magnetic shear and 
high current density (Thompson et al. 2000); or a persistent
vibration of the star (Fatuzzo \& Melia 1994). 

Measurement of the burst rise time $t_r$ provides a discriminant 
between these possibilities.  We find that $t_r$ is characteristically much
shorter than the total duration, as defined by either ${\rm T}_{90}$
or $\tau_{90}$. Moreover, the distribution of $t_r/{\rm T}_{90}$ is broad
(both for single-pulsed bursts and multi-peaked events),
which suggests that $t_r$ is not directly connected to cooling.
For example, the X-ray luminosity of a trapped fireball in local
thermodynamic equilibrium is proportional to its surface area, and
is a weak function of its internal temperature (Thompson \& Duncan 1995).
The characteristic radiative timescales are, in general, very short
at the high spectral intensity of an SGR burst. 
Thus, the rise of the X-ray flux plausibly represents
the initial injection of energy, although strictly $t_r$ only sets an 
upper bound to the injection timescale.

While the overall shape of T$_{90}$ (both single and multi-peaked) and 
the $\tau$$_{90}$ distributions of 
both sources are quite similar (relatively consistent Gaussian
widths), the peaks of both distributions for SGR 1900+14 occur at shorter
durations compared to those for SGR 1806$-$20. 
This systematic difference in the burst durations probably results
from some differing intrinsic property of the sources, such as the
strength of the magnetic field, or the size of the active region.

We find a power law correlation between the total burst counts (fluence) 
and the duration of SGR bursts with a power law index around 1. Similar 
behavior was noted for earthquakes by Gutenberg \& Richter (1956).
Within the context of earthquake mechanics, one way of defining duration
(also known as ``bracketed duration" [Bolt 1973]) is the time between the 
first and last 5\% excesses of g (gravitational acceleration on Earth) by 
the threshold acceleration of strong ground shaking. Recently, Lay \& Wallace 
(1995) presented the power law correlation between the seismic moment 
($\propto$ energy) and duration of 122 earthquakes with energies between 
3.5 $\times$ 10$^{23}$ erg and 2.8 $\times$ 10$^{26}$ erg. The power law 
fit to these events yields an index of 3.03.  

An equally important constraint on the injection and cooling mechanisms
comes from the anti-correlation between the hardness and fluence of the
SGR bursts.  Although very significant for SGR 
1806-20, this anti-correlation is much milder than that expected for 
black-body emission from a region of constant area; indeed,
the trend of increasing hardness with lower fluence 
is opposite to that expected for constant area emission.  
Two basic types of radiative mechanism could reproduce this trend.
First, the emitting plasma could be in local thermodynamic equilibrium, 
which requires that its size (radiative area) should decrease at lower
fluences.  An alternative possibility is that the spectral intensity 
of the radiation field sits below that of a black body, and that the
temperature of the emitting plasma is buffered within a narrow range.
We consider each of these possibilities in turn.

An SGR burst can be parameterized by the rate of injection of 
energy into the magnetosphere, $L_{\rm inj}$, and the volume $V$
of the injection region.  When $L_{\rm inj}$ is large and $V$ is
small enough, it is not possible to maintain a steady balance
between heating and radiative cooling.  The deposited energy is
locked onto closed magnetic field lines of the neutron star, in
a ``trapped fireball'' composed of photons and electron-positron pairs
(TD95).  This kind of event will tend to have a soft spectrum, because 
the injected energy has thermalized, and the plasma remains in LTE
very close to its photosphere.  The rise time is comparable to the
time over which energy is initially injected, but the decay is
limited by the rate of cooling through a thin radiative surface layer,
which contracts toward the center of the fireball.  The declining light
curve of the 27 August 1998 giant flare can be accurately fit by such
a model (Feroci et al. 2001;  Thompson \& Duncan 2001).  

A second burst from SGR 1900+14 on 29 August 1998 has been interpreted
in this trapped fireball model (Ibrahim et al. 2000).  The main 29 August burst
had a much shorter duration than the giant flare ($\sim 3.5$ s versus
$\sim 400$ s).  This bright component was followed by a much fainter 
pulsating tail (extending out to $\sim 1000$ s) which
provides direct evidence for heating and compression of the neutron
star surface by the fireball.  In this model, the short duration and 
high luminosity of the bright component require that the trapped fireball
had an approximately planar geometry, as would be expected if the energy
were released along an extended fault (Ibrahim et al. 2000).  

As our results show, most single-pulsed bursts from SGR 1900+14 have a 
much shorter duration (40 times smaller) than the bright component of 
the August 29 burst.  It remains unclear, therefore, whether
the trapped fireball model also applies to these much more frequent
events.  If it does apply, then the narrowness of the ${\rm T}_{90}$
distribution (compared with the wide range of measured fluences) also
requires a planar geometry, because the cooling time is determined
by the smallest dimension of the fireball.

The radiative mechanism is somewhat different if the injection luminosity
$L_{\rm inj}$ lies below a critical value of $\sim 10^{42}\,(V^{1/3}/
10~{\rm km})$ ergs s$^{-1}$ (assuming a spherical geometry;  
Thompson \& Duncan 2001).
When the compactness is that low,  it is possible to maintain a steady
balance between the heating of a corona of electron-positron
pairs, and radiative diffusion out of the corona (which one deduces must
be optically thick to scattering).  The central temperature of the corona 
is buffered in the range $\sim 20-40$ keV, and remains higher at
{\it lower} luminosities -- just the trend observed for SGR 1806-20 (Fig. 10).
If this radiative model applies to the majority of single-pulsed
events, then the constraints on the geometry of the active region are
weaker, and one must consider alternative mechanisms for
storing the injected energy, such as a persistent current driven
by shearing motions in the neutron star crust (Thompson et al. 2000).

\acknowledgments

E.G. is grateful to James C. Pechmann for useful discussions on earthquakes.
We acknowledge support from NASA grants NAG5-6021, NAG5-7785(E.G); 
the LTSA grant NAG 5-9350 (P.M.W. \& C.K.);
NASA grant NAG5-3100 and Alfred P. Sloan foundation (C.T.);
the Texas Advanced Research Project grant ARP-028 and 
NASA grant NAG5-8381 (R.C.D.). Several of the ideas presented in this work
originated at the ITP funded by NSF grant PHY99-07949.

\newpage

\begin{figure}
\plotone{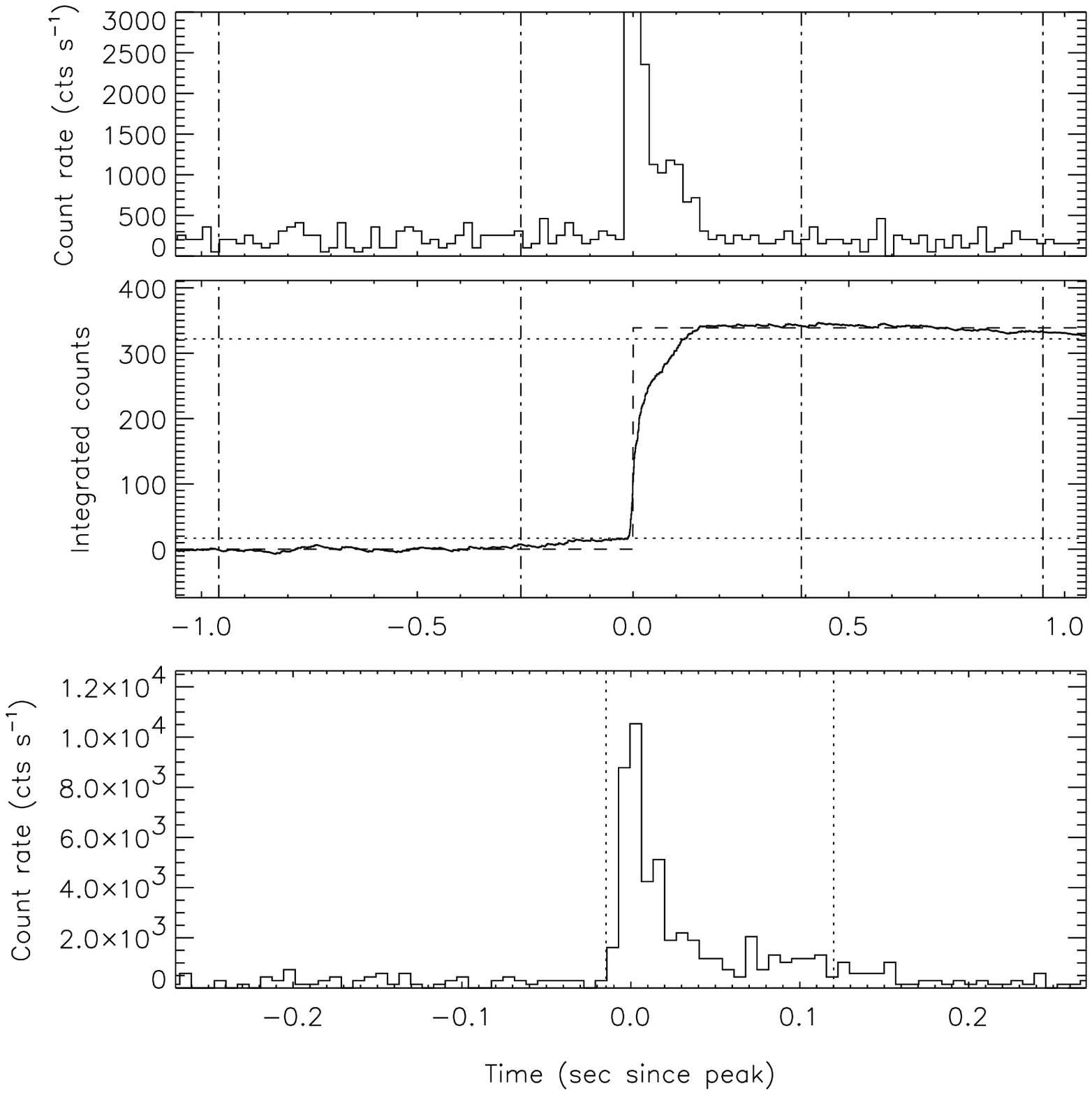}
\caption{({\it top panel}) Light curve of a SGR 1900+14 burst (2$-$60 keV) 
accumulated with 20 ms time resolution. The vertical dot-dash lines before and
after the burst are the boundaries of the selected pre- and post-background
intervals. ({\it middle panel}) The solid line shows the cumulative counts, 
with the background
contribution (as fit by a first-order polynomial) subtracted off.  The 
dashed line shows a step-function fit to the burst counts.
The lower horizontal dotted line represents the 5\% and upper horizontal 
dotted shows the 95\% count fluence level. Note that the top and the middle
panels are in the same time scale. ({\it bottom panel}) 
Time profile of the same burst
with 7 ms time resolution and zoomed in near the time of the burst. 
The vertical dotted lines
are the start and the end times of its T$_{90}$ duration.} 
\end{figure}

\newpage

\begin{figure}
\plotone{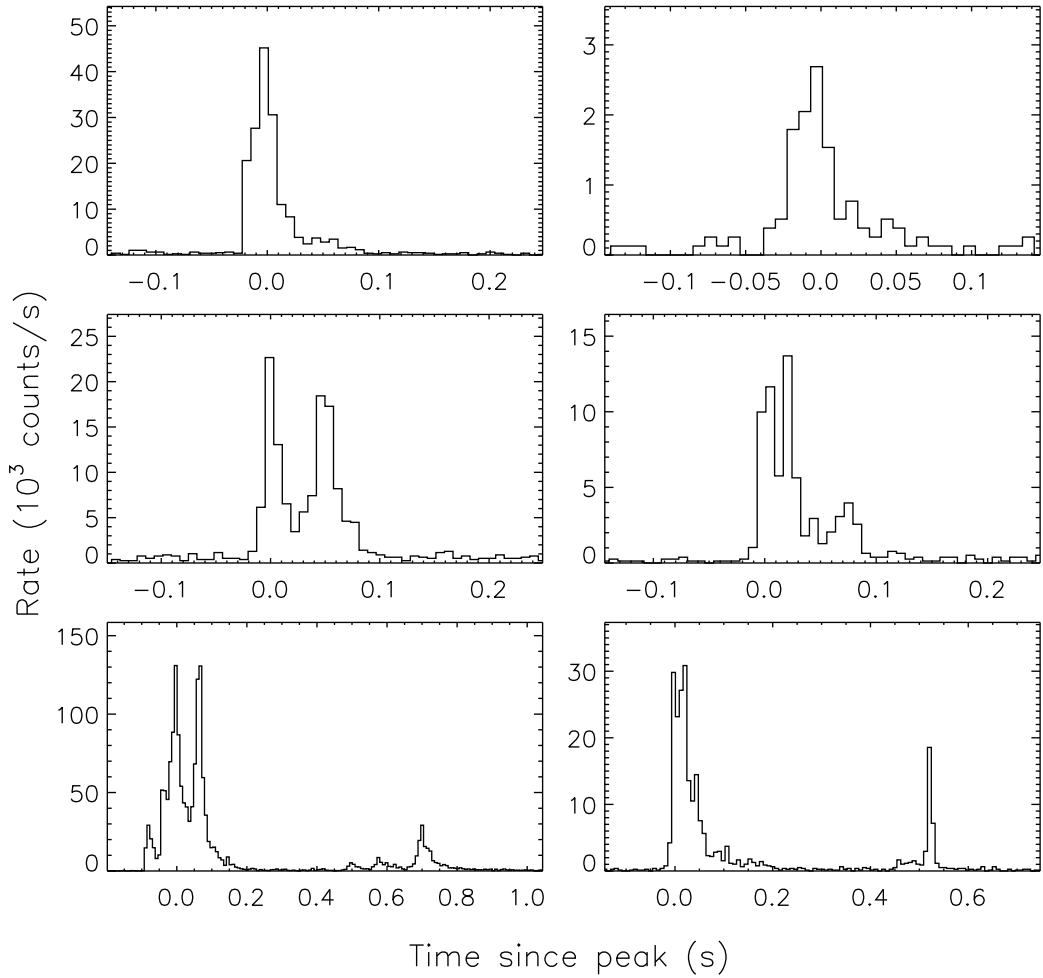}
\caption{Time profiles of some SGR 1900+14 and SGR 1806$-$20 bursts (on 7 ms
time scale) to illustrate single pulse bursts (top plots) and multi-peaked 
bursts (middle and bottom plots).}
\end{figure}

\newpage

\begin{figure}
\plotone{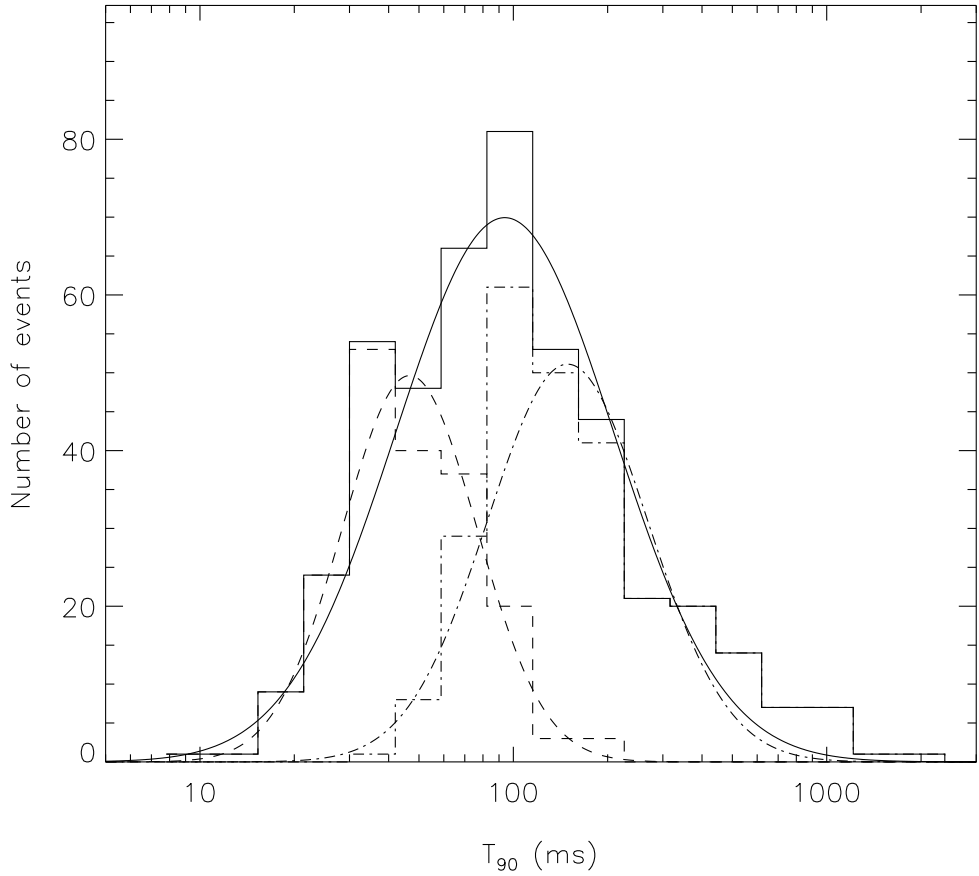}
\caption{Distributions of the T$_{90}$-durations for all (solid), single pulse 
(dashed) and multi-peaked (dash-dot) SGR 1900+14 bursts. 
The solid, dashed and dash-dot curves are obtained by fitting 
a log-Gaussian model to each set. Sets peak at 93.4 ms, 46.7 ms and 148.9 ms, 
respectively.}
\end{figure}

\newpage

\begin{figure}
\plotone{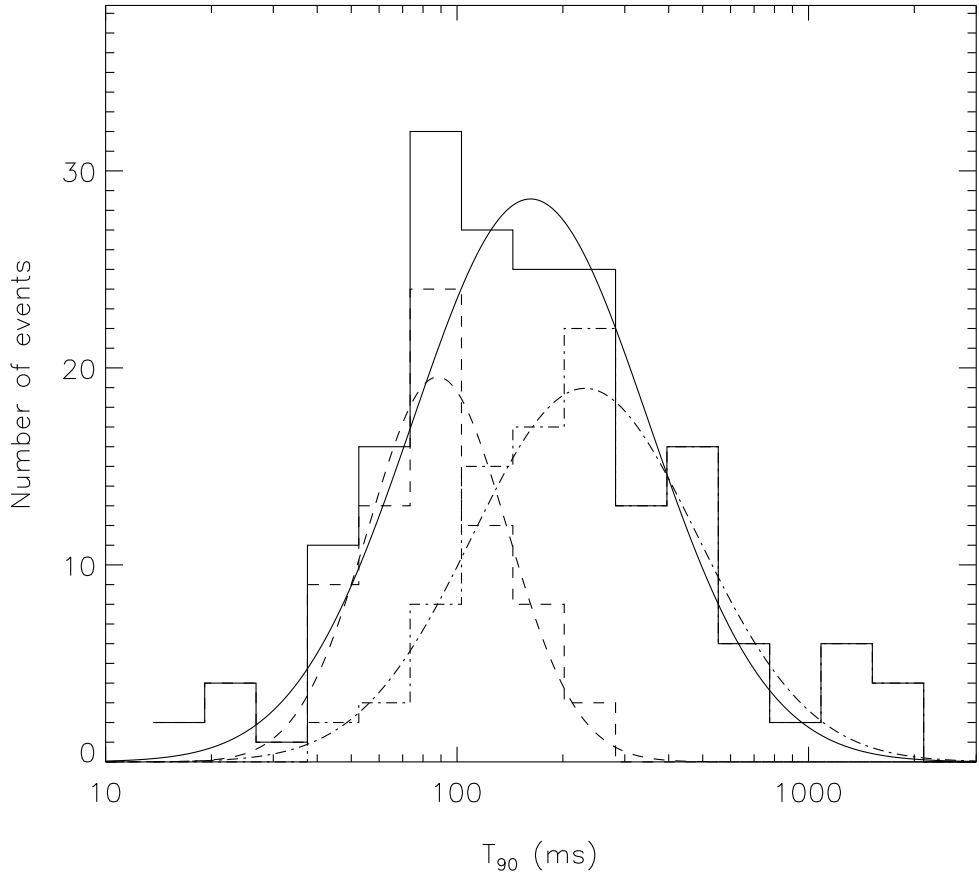}
\caption{Distributions of the T$_{90}$-durations for all (solid), single pulse 
(dashed) and multi-peaked (dash-dot) SGR 1806$-$20 bursts. 
The solid, dashed and dash-dot curves are obtained by fitting 
a log-Gaussian model to each set. Sets peak at 161.8 ms, 88.1 ms and 229.9 ms, 
respectively.} 
\end{figure}

\newpage

\begin{figure}
\plotone{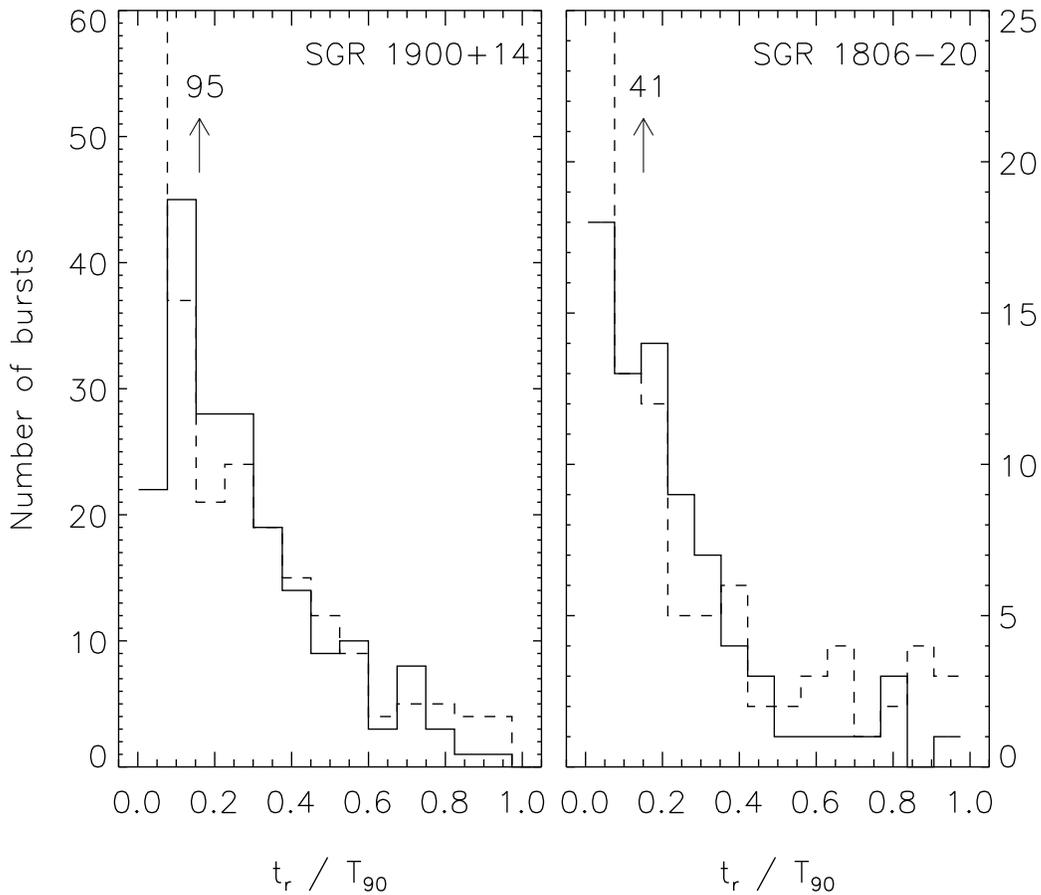}
\caption{Distributions of the ratios of rise times, t$_{r}$ to T$_{90}$ 
durations for single pulse bursts (solid) and multi-peaked bursts (dashed)
from SGR 1900+14 (left) and SGR 1806$-$20 (right). Note that the number of
multi-peaked events with the lowest t$_{r}$/T$_{90}$ ratios is artificially 
large due to the longer (compared to t$_{r}$) durations of these bursts 
(see Figure 2, middle and bottom plots).}
\end{figure}

\newpage

\begin{figure}
\plotone{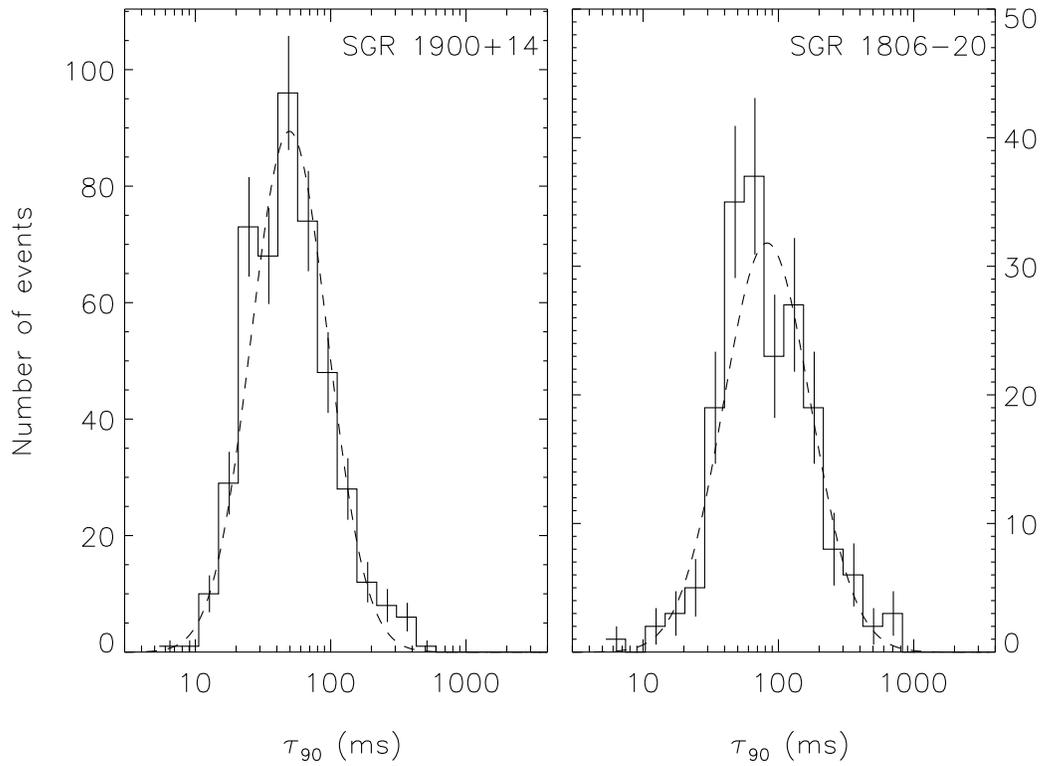}
\caption{Distributions of the $\tau$$_{90}$ emission times for SGR 1900+14 
(left) and SGR 1806$-$20 (right) bursts. The dashed curves are obtained by 
fitting a
log-Gaussian model to each set, with peaks at 49.6 ms and 82.3 ms, 
respectively.}
\end{figure}

\newpage

\begin{figure}
\plotone{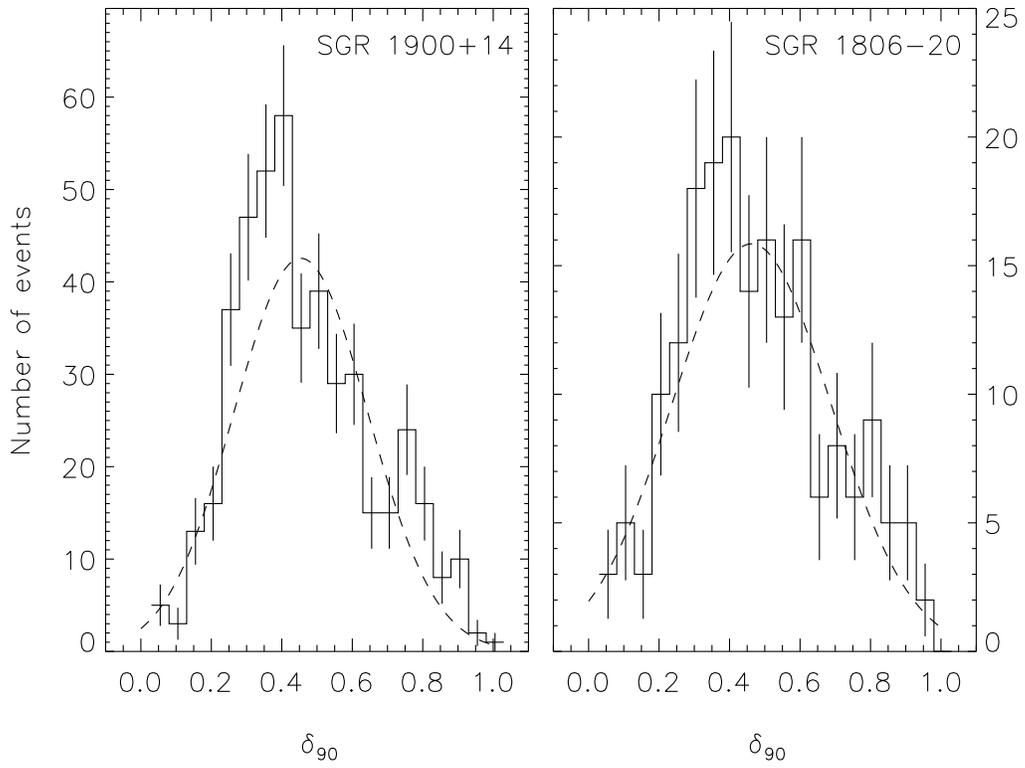}
\caption{Distributions of the duty cycles, $\delta$$_{90}$, for SGR 1900+14 
(left) and SGR 1806$-$20 (right) bursts. The Gaussian fits peak at 0.45 and 0.46,
respectively.}
\end{figure}

\newpage

\begin{figure}
\plotone{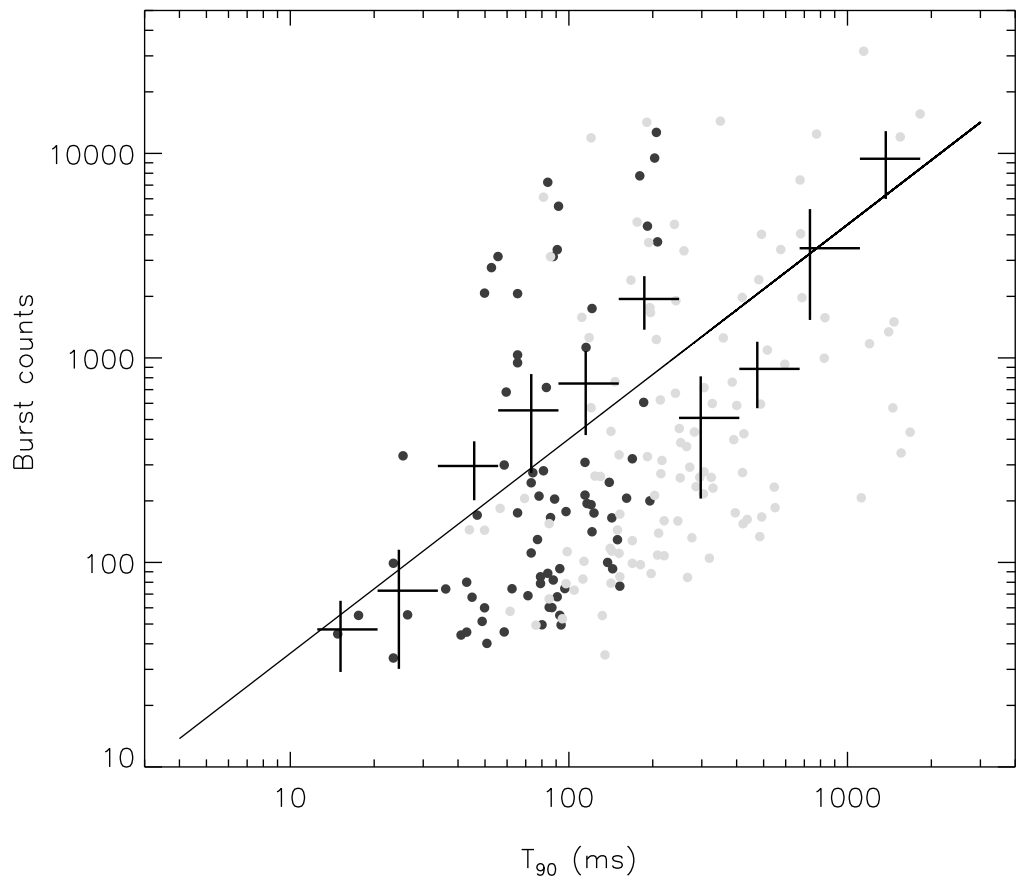}
\caption{Plot of the total counts vs T$_{90}$-durations for 190 SGR 1806$-$
20 bursts which shows a correlation ($\rho$ = 0.91, P=3.4 $\times$ 10$^{-4}$). 
The solid line is obtained by fitting a power law model to the crosses (which
are explained in the text). The best fit power law index is 1.05 $\pm$ 0.16.
The circles are the individual measurements for single pulse (dark) and 
multi-peaked (gray) bursts.}
\end{figure}

\newpage

\begin{figure}
\plotone{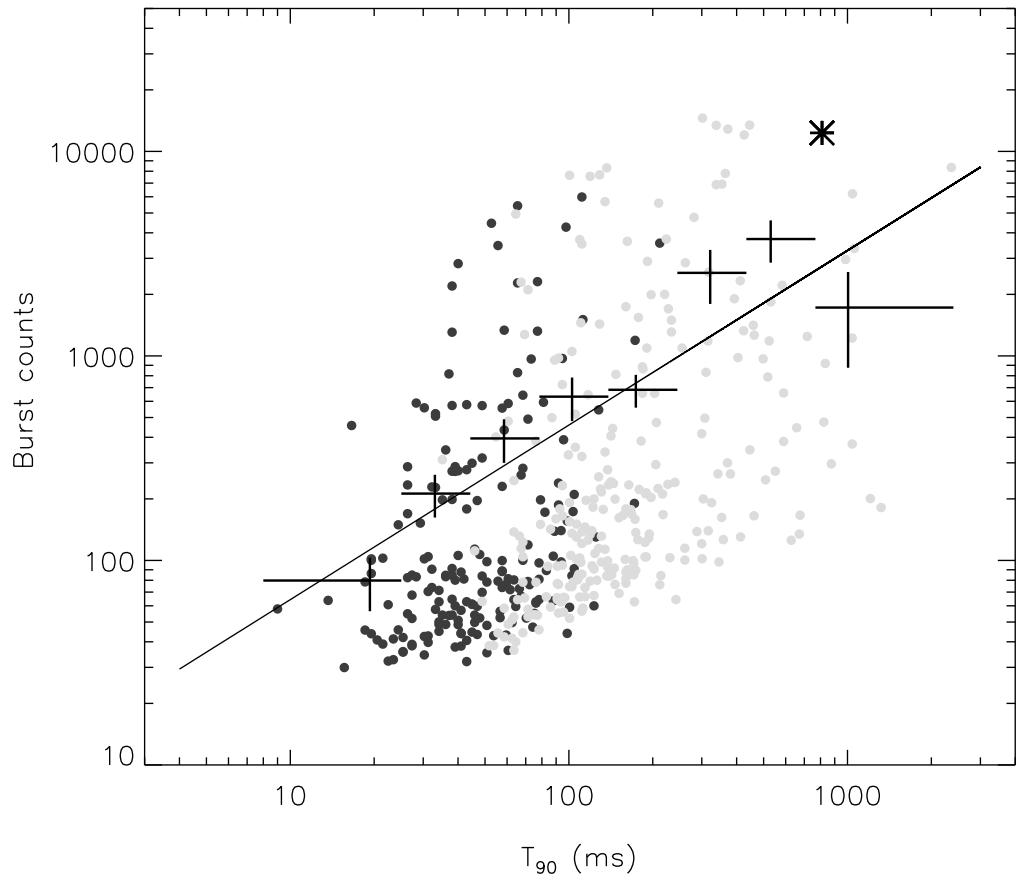}
\caption{Plot of the total counts vs the T$_{90}$-durations for 455 SGR 
1900+14 bursts which shows a correlation ($\rho$ = 0.89 and P = 8.6 
$\times$ 10$^{-4}$). A power law fit to the crosses (solid line) yields an 
index of 0.91 $\pm$ 0.07. The asterisk indicates where August 29 precursor
would be seen. The circles represent the measured values for single pulse 
(dark) and multi-peaked (gray) bursts.}
\end{figure}

\newpage

\begin{figure}
\plotone{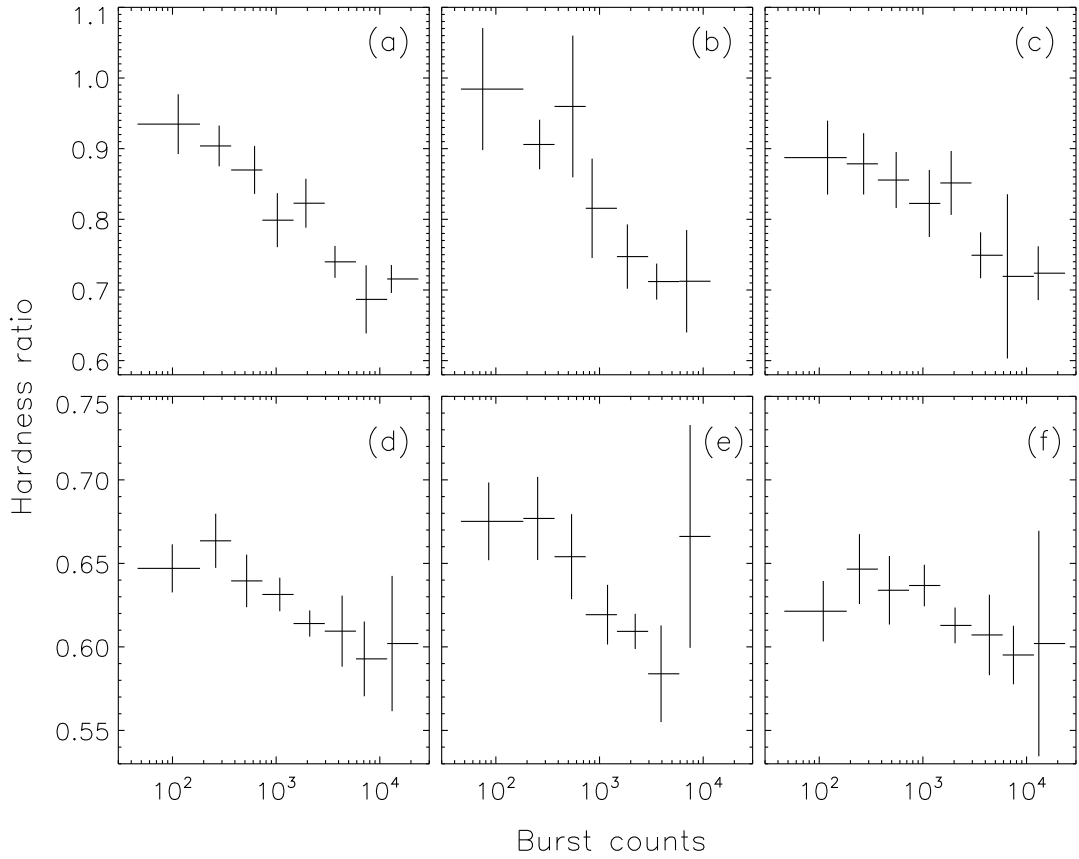}
\caption{Plot of hardness ratios vs total counts for (a) all, (b) single 
pulse, (c) multi-peaked SGR 1806$-$20 bursts. Similarly, those for (d) all,
(e) single pulse, (f) multi-peaked SGR 1900+14 events.}
\end{figure}

\newpage

\begin{figure}
\plotone{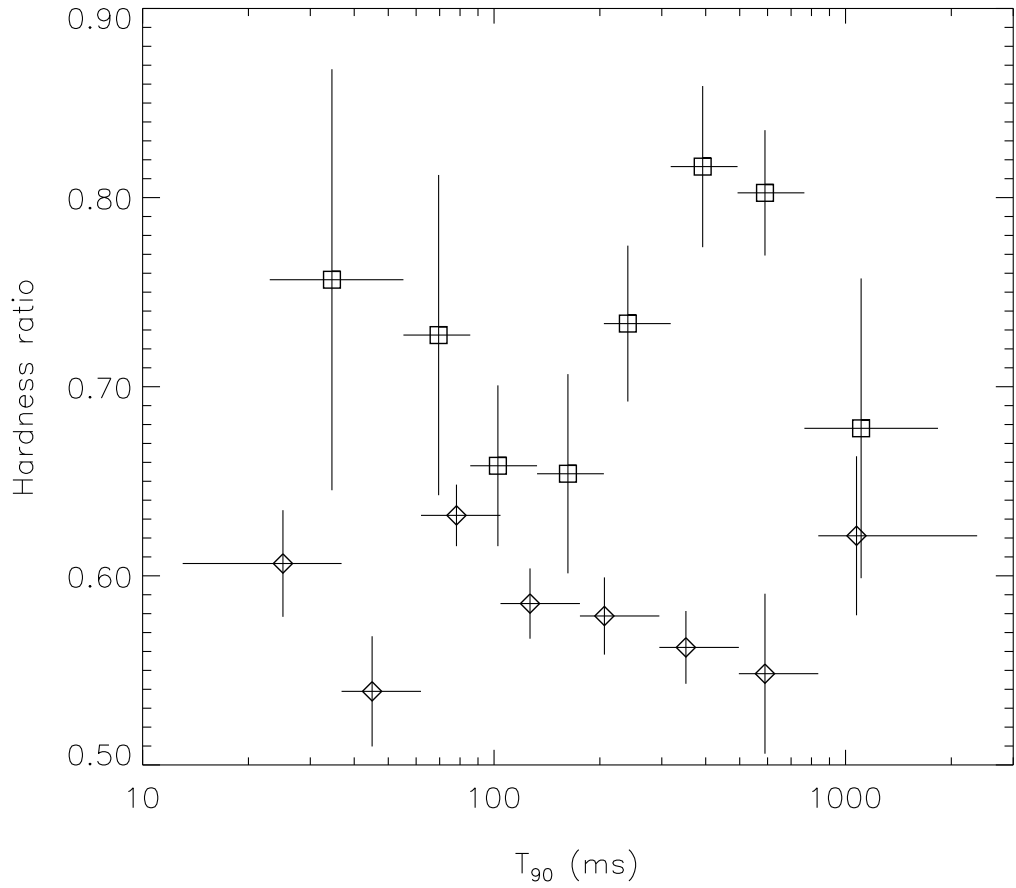}
\caption{Plot of hardness ratios vs T$_{90}$ durations for SGR 1806$-$20 bursts
(squares) and for SGR 1900+14 (diamonds).}
\end{figure}

\end{document}